\documentclass[published]{JHEP3} 
\accepted{July 16, 2002}
\received{June 3, 2002}
\revised{July 4, 2002}
\JHEP{07(2002)039}

\title{Reality, measurement and locality in quantum field theory}

\author{Daniele Tommasini\\ 
Departamento de F\'\i sica Aplicada,
\'Area de F\'\i sica Te\'orica, Universidad de Vigo\\ 
32004 Ourense, Spain\\ 
E-mail: \email{daniele@uvigo.es}}

\abstract{It is currently believed that the local causality of Quantum
Field Theory (QFT) is destroyed by the measurement process. This
belief is also based on the Einstein-Podolsky-Rosen (EPR) paradox and
on the so-called Bell's theorem, that are thought to prove the
existence of a mysterious, instantaneous action between distant
measurements. However, I have shown recently that the EPR argument is
removed, in an interpretation-independent way, by taking into account
the fact that the Standard Model of Particle Physics prevents the
production of entangled states with a definite number of
particles. This result is used here to argue in favor of a statistical
interpretation of QFT and to show that it allows for a full
reconciliation with locality and causality. Within such an
interpretation, as Ballentine and Jarret pointed out long ago, Bell's
theorem does not demonstrate any nonlocality.}

\begin{document}

\section{Introduction} 

Quantum Field Theory (QFT) QFT has been argued to be the unique
reasonable realization of a Relativistic Quantum
Mechanics~\cite{WeinbookI}, and it can be thought to be the ``low
energy" limit of any possible Theory of Everything. In fact, an
extremely successful description of Particle Physics, valid at least
up to energies $\sim100$\,GeV, has been obtained with the Weinberg -
Salam Standard Model~\cite{WeinbookII}, which is based on the
hypothesis that all the known interactions (except gravity) can be
derived from a \emph{local} symmetry
SU(3)$\times$SU(2)$\times$SU(1). The field equations define a
perfectly Lorentz-invariant, causal and local
theory~\cite{WeinbookI,WeinbookII}. Of course, these characteristics
are most welcome in a relativistic world. However, they are usually
thought to be broken by the process of
measurement~\cite{measrel,measloca,Zeilinger,Laloe,Scarani}.

In fact, the measurement process is usually described by a ``collapse"
of the state vector of the system, which after measurement is
projected onto an ``out" state, an eigenstate of the observed
quantities (e.g.\ the momenta and helicities of the observed
particles). This collapse postulate works to obtain correct
theoretical predictions, however it implies that the measurement
process itself is a sort of discontinuous process that cannot be
described with the smooth, linear evolution that applies to all known
interactions (this is the so-called ``measurement problem", see
refs.~\cite{Ballentine70,Laloe} for reviews showing different points
of view). Moreover, the ``collapse" is described as a global effect,
involving at the same time all the space.

Is such a supposed nonlocality a real physical property of any quantum
theory, and possibly of Nature (as it is usually supposed today)? Or
is that merely a problem of the global collapse description, i.e.\ of
the ``interpretation" of the theory? Is it possible to avoid the
collapse and the measurement problem, and describe the measurement
process with the same local QFT laws that describe particle
interactions?

\section{Reality and completeness: the EPR paradox} 

In their famous 1935 paper~\cite{EPR}, EPR provided two important
concepts that have greatly influenced the subsequent research in the
foundations of the quantum theory.
{\renewcommand{\theenumi}{\Roman{enumi}}
\begin{enumerate}   
\item A ``condition of completeness" for any acceptable theory: {\it
``every element of the physical reality must have a counterpart in
the physical theory"}~\cite{EPR}.

\item A definition of the objective ``physical reality": \emph{``If,
without in any way disturbing a system we can predict with
certainty (i.e.\ with probability equal to unity) the value of a
physical quantity, then there is an element of physical reality
corresponding to this physical quantity"}~\cite{EPR}.
\end{enumerate}}

These concepts were aimed at dealing with the main difficulty of
Quantum Mechanics (QM): the particles and the physical quantities are
usually not definite (i.e.\ they have no ``reality") until a
measurement is performed.\footnote{This is just a generalization of
the so-called wave-particle duality.} However, EPR pointed out that
even ordinary QM allows for some elements of physical reality: if a
system is prepared in an eigenstate of a given observable, we can
predict with certainty that the result of the measurement of that
observable will be the corresponding eigenvalue: there is then an
element of objective physical reality corresponding to that
observable.

In Classical Physics, all the physical quantities have a definite
value in a given system at a given time (although in the case of
macroscopic systems we usually do not know such a value and can only
use a statistical description). In the usual ``orthodox"
interpretation of QM, however, it is supposed that the state vector
completely describes the state of a system, and this does not allow
for a certain prediction of the results of the measurements of two
noncommuting observables, such as the position and the momentum. Given
a state, unavoidably there are some observables (heuristically, a
``half" of the set of the observables) that do not have a reality, and
their measurement in an ensemble of copies of the system prepared in
this state will show a nonvanishing dispersion. Are these
non-diagonalized physical quantities really undefined on the single
copy of the system that is under consideration? Or is this uncertainty
merely a consequence of an unavoidable lack of knowledge, as in the
classical, statistical description of the macroscopic systems? In the
latter case, the QM description should be completed by introducing
some new, ``hidden" variables, possibly allowing for an underlying
determinism.

The brilliant argument developed by EPR was aimed at resolving this
dilemma. In fact, they invented a thought-experiment for which QM
itself predicted that two incompatible observables (position and
momentum in their original formulation; two noncommuting components of
the spin in a version due to Bohm~\cite{Bohm}) were given a
simultaneous reality. They argued that this result was in
contradiction with the assumption that the wave function completely
described the physical reality.

Although all the present discussion can be easily generalized to EPR
experiments involving any kind of particles, hereafter I will only
consider the so-called ``EPR-Bohm"
thought-experiment~\cite{Bohm,pureprp}. Two spin $1/2$ particles, A
and B, are created in coincidence in a spin-singlet state, and are
detected by the detectors $O_A$ and $O_B$ in opposite directions. The
measurement of a given component $\vec S\cdot\vec a$ of the spin of
particle A (or of B) along a unit vector $\vec a$ can give the values
$\pm\hbar/2$, each with probability $1/2$. However, if $\vec
S\cdot\vec a$ is measured on A and found, say, equal to $+\hbar/2$,
not only does this give a physical reality to such a spin component on
A; in fact, momentum conservation allows for predicting with certainty
the value $-\hbar/2$ for the same spin component on B. EPR assumed
that \emph{the physical reality on B is independent of what is done
with A, which is spatially separated from the former} (this assumption
has been called \emph{Einstein's Locality}). Since a certain
prediction for the considered spin component on B was allowed
\emph{without in any way disturbing particle B}, therefore \emph{the
spin component $\vec S\cdot\vec a$ has a reality on B}. By repeating
this argument for any component of the spin, we deduce that \emph{all
the spin components ($S_x$, $S_y$, $S_z$) have a simultaneous physical
reality} on particle B.  But this contradicts ordinary QM as based on
the wave function, where only one component of the spin of a given
particle can be definite.

This result is a rigorous consequence of two assumptions, as Einstein
himself noticed in 1949~\cite{Einstein49}:

``The paradox forces us to relinquish one of the following two
assertions:"
\begin{enumerate} 
\item the description by means of the wave function is {\it
complete},

\item the physical realities of spatially separated objects are
independent of each other.
\end{enumerate}

The incompatibility of statements 1) and 2) has also been called
\emph{EPR theorem} (see e.g.\ refs.~\cite{Ballentine70,Peres,Laloe}).

Since ``Einstein's locality" assertion 2) was considered
unquestionable by EPR, they deduced that the wave function did not
provide a complete description of the state of a system. This was a
very powerful argument in favor of deterministic (and local) hidden
variable theories. In such theories, all the observables have a
definite value in the single system that is under consideration. The
dispersion of the probability distributions observed in the repetition
of the experiment on an ensemble of identically prepared systems is
merely a ``statistical mechanics" effect. Such a ``cryptodeterminism"
(see page 155 of ref.~\cite{Peres}) would explain the fact that the
measurement of a component of the spin of A apparently has a
deterministic effect on the distant measurement of the same component
of the spin of B: both results would actually be the deterministic
consequence of the common production of the two particles.

\section{The supposed proofs of nonlocality} 

In the mid sixties, Bell proved that deterministic hidden variable
theories were actually viable, but they could not reproduce all the
results of QM~\cite{Bell64}, unless they implied an instantaneous
action at a distance~\cite{Bell66}. He proposed a set of ``Bell's
inequalities" for the spin correlations in a realization of the
EPR-Bohm experiment, that were violated by QM and respected by any
local deterministic hidden variable theory. Since the actual
experiments~\cite{Aspect} confirmed the predictions of QM, local
determinism was ruled out. \emph{This result will be called hereafter
the ``original Bell's theorem".}

Therefore, it was deduced that QM was a complete theory, and EPR
Theorem forced to conclude that it had to be a ``nonlocal"
theory.\footnote{This conclusion also seemed to be suggested by the
very fact that it was actually possible to reproduce the QM results
using an explicitly nonlocal deterministic theory that was found by
Bohm~\cite{Bohm52}.} Hereafter, I will call this argument the
``EPR+Bell" proof of nonlocality, since it is based on both EPR and
Bell's theorems.\footnote{Many of the existing supposed proofs of
nonlocality are reelaborations or variations of this argument.}

Moreover, in the last several years there has been a proliferation
of \emph{``generalized Bell's Theorems",} claiming that the
observed violation of  Bell's inequalities was sufficient in
itself to prove the existence of an instantaneous influence
between distant measurements.

As a result of these two proofs, there has been increasing agreement
within quantum physics experts\footnote{There have been remarkable
exceptions, however, mainly represented by researchers resisting
giving up local hidden variable theories. All of them would deserve a
citation, but I can only give space to the argument published in 1987
by Ballentine and Jarret~\cite{BaJa87}, whose main conclusions are
still valid today, although they are almost universally ignored.} on
the conclusion that Nature is EPR paradoxical. ``Quantum nonlocality"
is considered an experimental evidence, and it is really believed that
the two distant measurements in the EPR experiments do actually
influence each other instantaneously. The ``speed of quantum
information" has even been ``measured" to be greater than
$1.5\times10^{4}c$~\cite{Scarani}.

Einstein could never accept the existence of such a ``spooky action at
a distance", which is incompatible with special relativity. In fact,
this supposed ``quantum nonlocality" is the main origin of the
widespread belief that the Quantum Theory is incompatible with Special
Relativity (see e.g.\ refs.~\cite{measrel,Scarani}),\footnote{As
Ballentine and Jarrett remarked, ``such a contradiction, if indeed one
exists, between two fundamental and exceedingly well verified theories
would constitute a major crisis in theoretical physics. One may wonder
why so few physicists are apparently concerned (or even aware of) such
a crisis"~\cite{BaJa87}. Surprisingly, the same consideration can be
repeated today, to an even greater extent.} although it is recognized
that the EPR correlations do not allow for superluminal signaling,
e.g.\ they cannot be used to synchronize clocks (see ref.~\cite{Laloe}
for a review).

In fact, any kind of ``nonlocality" or instantaneous ``distant
influence" is unacceptable in a relativistic world: due to the
relativity of the simultaneity, suitable observers would describe this
influence (whatever it is) as an effect of the future on the
past~\cite{BaJa87}. Fortunately for science, \emph{there is a way to
completely reconcile the quantum theory with locality, causality and
special relativity}. This solution is quite natural and is based
merely on known physics.

\section{Photons uncertainty, reality and the EPR paradox} 

The EPR argument, as described above, and (as far as I know) all the
subsequent treatments of the EPR paradox, have assumed that it was
actually possible to prepare a system of two entangled particles.  In
fact, for the EPR argument it is crucial that the measurement on A
implies a \emph{certain} prediction for B without disturbing B: thus
different noncommuting observables on B are forced to have a physical
reality.\footnote{EPR explicitly asked for a unit probability for this
prediction, but we can replace their requirement with the weaker one:
\emph{with probability $1-\epsilon$, where $\epsilon$ may be made
arbitrarily small}~\cite{Ballentine70}. The following discussion
remains true.}

However, I have recently proved that this assumption is not
correct~\cite{pureprp}. In fact, the Standard Model of Particle
Physics predicts that it is not possible to produce a state having a
definite particle content: \emph{given the process that produces A and
B alone, QFT theory predicts a nonvanishing and finite probability for
the creation of A and B plus additional photons}~\cite{pureprp}. In
other words, given the initial system that is used for the production
of A and B, \emph{the rate for the processes involving additional
photons is a given finite number that cannot be made arbitrarily
small.}

As a consequence, I argued that it is \emph{never} possible to make
a certain prediction of the spin state of B by merely measuring a
spin component of particle A in an EPR experiment. Energy,
momentum and angular momentum conservation do not hold for the
(sub)system made of the two particles A and B, since additional
photons can carry such conserved quantities. By detecting A, it is
not even possible to predict if particle B will actually be caught
in the opposite direction~\cite{pureprp}.

The conclusion may seem surprising: \emph{QFT satisfies the EPR
criterion of completeness without the need for hidden variables and
without necessarily violating Einstein's condition of locality for the
physical reality}~\cite{pureprp}. In other words, \emph{the EPR
argument is removed because of the fact that QFT allows for less EPR
reality than was believed.} This solution, based on the greater
uncertainty of QFT, is \emph{completely opposite} to the local
determinism solution based on hidden variables.  Incidentally, since
it is impossible to \emph{prepare} a state with a definite number of
photons, and since such an uncertainty for any given process cannot be
made arbitrarily small, we may even argue that it is not possible to
give a physical reality (not even locally) to any observable, except
to the charges and masses (that are the invariants of the
theory).\footnote{In QFT, the usual assumption that the state of a
system can be \emph{prepared} by measuring a Complete Set of Commuting
Observables is also incorrect, e.g.\ it is impossible to determine the
occupation numbers of the photon states arising from a given
process. Of course, the usual Fock space basis of QFT, diagonalizing
all the occupation numbers, can still be used as a mathematical tool.}

Therefore, we no longer have to decide between statements 1) and 2) in
EPR theorem: QFT remains an EPR-complete theory and yet it does not
satisfy condition 1), that implicitly assumed the entangled two
particle state vector. Therefore QFT can satisfy 2) in EPR theorem. In
particular, \emph{the EPR+Bell proof of nonlocality is removed.}

This result is particularly significant since it relies on only
well-established physics and it does not depend on any particular
interpretation of the theory. However, it is clear that a decision
about the locality of QFT can be made only after examining the
description of the measurement process. If (and only if) this can be
done in a local way, without postulating any global collapse of the
state vector, then QFT will be automatically a causal and local
theory.\footnote{Incidentally, it is obvious that the orthodox
interpretation will be nonlocal, since it postulates such a global
collapse. I do not need Bell's Theorems to believe that.} We will now
look for an interpretation that executes this program.

\section{The interpretation of QFT}

According to the previous discussion, QFT can be considered an
EPR-complete theory. If we do not introduce any hidden variables or
new physics beyond QFT, we are left with two possibilities for an
interpretation:
{\renewcommand{\theenumi}{\Roman{enumi}}  
\begin{enumerate} 
\item The state vector applies to the single system. Now, since the state
vector of a free system evolves deterministically in Quantum Mechanics
and in QFT, the joint state of the measuring apparatus (including all
the ``environment" which interacts with it) and the object system
after the measurement has to be determined by that before the
measurement. In particular, the position of the pointer, i.e.\ the
result of the measurement, has to be the deterministic consequence of
the initial conditions, and the only possible source of indeterminism
is the unavoidable statistical ignorance of the state of the
environmental variables. Since the QFT laws are local, the world would
be intrinsically deterministic and local. This possibility is
logically consistent, but it seems to be unable to provide a
satisfactory solution to the ``measurement problem" (see below).

The ``orthodox" interpretations~\cite{Laloe,Peres} introduce the
collapse postulate in order to reconcile the assumption that the
state vector applies to the single system with the fact that the
measurement gives sharp results. The measurement process is then
considered a magical process, different from the physical
interactions that are well described by QFT. I think that this ad
hoc assumption is very unnatural in QFT; it also contradicts the
experience in particle physics detectors, that shows that the
measurement process is determined by the same strong and
electroweak interactions that are described by QFT. Moreover, the
collapse of the state vector has to simultaneously involve all the
space and explicitly violates locality and Lorentz symmetry. This
is unacceptable in a relativistic world, as discussed above.
Paradoxically, in spite of all these contradictions and
absurdities, orthodox interpretations are currently the most
common choice. Up to now, this may have been justified precisely
by the pressure of both EPR and Bell's theorems taken together.

In non-relativistic Quantum Mechanics, a more reasonable framework to
assign a state vector to the single system is based on a stochastic
modification of Schr\"odinger's equation~\cite{stochastic}. Here, I
will not discuss these kinds of interpretations since for the moment
they have not been implemented to take into account Special Relativity
in a completely consistent way. Moreover, they go somewhat beyond the
purposes of the present paper since they suppose the introduction of
new physics beyond QFT.

\item The remaining possibility for interpreting QFT without
introducing new physics is to assume that the state vector does
not describe the single system (which I will also call event), but
only describes an ensemble of identically prepared copies of the
system (more precisely, statistical operators should be used,
since photons uncertainty prevents the preparation of a pure state
as discussed below). This minimal statistical interpretation is
clearly more economical than orthodox ones that use the
unnecessary assumption that the state vector also describes the
single systems. Moreover, it is very natural in QFT: for instance,
if we consider a single particle that can decay, the theory is not
able to say when it will actually decay, which direction the
particles that are produced in the decay process will take, and
which of the possible decay channels will be chosen.\footnote{Note
that the result summarized in the previous section imply that
there are \emph{always} an infinite number of possible decay
channels, involving an arbitrary number of additional photons; of
course, usually there are also different channels involving other
particles, such as neutrinos or different charged particles (for
sufficient mass of the decaying particle).} Similarly, in a Young
double slit experiment, we are not able to predict anything about
the single particle that will hit the screen in a random position.

As a matter of fact, the photons uncertainty itself implies that
\emph{QFT does not make any prediction on the single event} (i.e.\ on
the single copy of the system), with the exception of the charges and
masses of the particles that will possibly appear.  However, \emph{QFT
predicts probability rates and cross sections that can be compared
with the frequencies of the results for the repetition of an
experiment on a statistical ensemble of equally-prepared copies of the
considered system.} Therefore, it is natural (or at least
conservative) to assume that the state vector only describes the
ensemble, and not the single copy of the system (that we do not
describe at all).
\end{enumerate}}

Of course, very convincing arguments for a statistical
interpretations of  the quantum theory have already been provided
in literature, see e.g.\ refs.~\cite{Ballentine70,Belinfante,BJ}
and references therein. The most important is the fact that it
naturally removes the measurement problem. Let me briefly
summarize how this actually occurs.

For simplicity's sake, I will consider the measurement of an
observable ${\cal A}$ in a two dimensional space of states. Let
$\alpha_1$ and $\alpha_2$ be the (different) eigenvalues of ${\cal
A}$, and let $\vert \alpha_1\rangle$, $\vert \alpha_2\rangle$ the
corresponding eigenstates. If the (normalized) state of the object
system is $\vert\psi\rangle = c_1\vert \alpha_1\rangle+c_2\vert
\alpha_2\rangle$, then the measurement of ${\cal A}$ will give either
the result $\alpha_1$ with probability $\vert c_1\vert^2$ or the
result $\alpha_2$ with probability $\vert c_2\vert^2$.  After the
measurement, the usual collapse postulate implies that the state is
reduced to one of the eigenvectors of ${\cal A}$, depending on the
result of the measurement: for instance, if the result is $\alpha_1$,
the state after the measurement is reduced to $\vert
\alpha_1\rangle$. The problem is that this postulate is incompatible
with the Schr\"odinger equation (or the field equations) that imply a
smooth and lineal evolution. In fact, let $\vert\Phi\rangle$ be the
intial state of the measuring apparatus (that belongs to a different
Hilbert space than that of the object system), and let $\Phi_i$ be its
states when the results $\alpha_i$ are obtained, for $i=1,2$
respectively (let these vectors also describe the so-called
``environment"). This means that if the initial state of the object
system was one of the eigenvectors of ${\cal A}$, say $\vert
\alpha_1\rangle$, then the corresponding initial state for the
composite system made of our object + measurement device would be the
tensor product $\vert \alpha_1\rangle\vert\Phi\rangle$, and the state
after the measurement would be the tensor product $\vert
\alpha_1\rangle\vert\Phi_1\rangle$. Similarly, $\vert
\alpha_2\rangle\vert\Phi\rangle\to\vert
\alpha_2\rangle\vert\Phi_2\rangle$ as a consequence of the
measurement. If the measurement process could be described by the same
linear evolution that applies to any physical process (the
Schr\"odinger equation or the field equations), when the object system
is not in an eigenstate of ${\cal A}$ the previous equations
necessarily imply the following result for the measurement
\begin{equation}
(c_1\vert \alpha_1\rangle+c_2\vert \alpha_2\rangle)
\vert\Phi\rangle\to c_1\vert \alpha_1\rangle\vert\Phi_1\rangle +
c_2\vert \alpha_2\rangle\vert\Phi_2\rangle\,. \label{eq:1}
\end{equation}
Now, if we assign a state to the individual system, as done in the
usual ``orthodox" interpretation of QM, we get a wrong result, since
after measurement the system will be put in one of the two
eigenvectors, not in the linear combination of eq.~(\ref{eq:1}) (a
measurement gives a definite result, not a superposition). This is the
so-called \emph{measurement problem.} The usual ``orthodox"
interpretation introduces the collapse of the state vector by hand,
and therefore requires a nonlinear evolution law for the measurement
process, which is different from the linear evolution that describes
any other process (in other words, within this interpretation the
measurement process cannot be described by the quantum theory
itself). In my opinion, any interpretation that does not give a better
answer to the measurement problem should simply be discarded, for
several good reasons: we know from half a century of detector building
that the particles are detected due to the usual electroweak and
strong interactions, that are described by the linear equations of the
Standard Model. Moreover, the collapse postulate is ultimately
responsible for the paradoxes (that we might better call
inconsistencies) of the quantum theory~\cite{Ballentine70,Laloe}. In
particular, it introduces a privileged reference frame, the
Laboratory; the global description of the collapse is responsible for
the supposed, paradoxical nonlocality of the theory, thus violating
Special Relativity.

How does the statistical interpretation avoid the ``measurement
problem"? It assumes that the state vector only describes a
statistical \emph{ensemble} ${\cal E}$ of identical copies of the
considered system, and it does not describe the single copy. The
state after the measurement \emph{is} the linear superposition of
eq.~(\ref{eq:1}). In fact, this state correctly describes the
probabilities $\vert c_1\vert^2$ and $\vert c_2\vert^2$  for
having obtained $\alpha_1$ or $\alpha_2$. Moreover, the single
event \emph{has} a definite result, although the theory does not
predict it. The events that have given the result $\alpha_1$ form
a subensemble ${\cal E}_1$ of ${\cal E}$, while those that gave
$\alpha_2$ correspond to a different subensemble ${\cal E}_2$
(with ${\cal E}={\cal E}_1U{\cal E}_2$). After the measurement,
the experimenter (or the measuring device) selects only the events
that have given a particular result, say $\alpha_1$, and this
corresponds to reducing the statistical ensemble to  ${\cal E}_1$.
After this selection of the events, the state vector is the one
that describes the new ensemble ${\cal E}_1$, which is $\vert
\alpha_1\rangle\vert\Phi_1\rangle$, as in the usual collapse
postulate (but here we have no paradox since we are not
associating such a state to the individual system and we did not
need a nonlinear evolution during the measurement process).

I recommend the excellent reviews~\cite{Ballentine70,Belinfante,BJ} for a more complete discussion
in terms of mixed states (which may be more fundamental than the
pure states). Note that in ref.~\cite{Ballentine70} the EPR
paradox led to the conclusion that the quantum theory was
incomplete and had to be completed: the single system had to have
precise values of anticommuting observables like momenta and
positions. Here, we do not need this assumption. QFT satisfies the
EPR condition of completeness, without introducing additional
variables. It is a strange kind of completeness: it allows for
less reality than ordinary QM. Thus QFT can only be used to
predict probabilities, statistical averages, correlations (that
can be computed as usual through Feynman diagrams). Although I
prefer to use the term ``statistical", this is essentially the
``correlation interpretation" described in ref.~\cite{Laloe} (see
also ref.~\cite{PeresFuchs}).

Incidentally, within this interpretation the world (i.e.\ the
single event) \emph{has} an objective consistency, independently of
the possible measurements (no need to rely on any mysterious
``quantum consciousness"). We merely cannot \emph{describe} it but
statistically.

As a result, the measurement process can be considered a physical
process involving quantum systems, that can be described by the
quantum theory itself. On the other hand, the viability of such a
description has been confirmed explicitly by recent studies,
relying for the moment on the nonrelativistic QM (see e.g.~\cite{Zeno} and references therein). Here, I will generalize this
conclusion by assuming that the measurement process may be
described precisely by the Standard Model itself, in agreement
with the actual experience of Particle Physicists. This implies
that {QFT with this statistical interpretation is a fully local
theory.} In fact, causality and locality are satisfied by the
Green functions of the theory (and then by the scattering matrix)~\cite{WeinbookI,WeinbookII}. In other words, \emph{there is no
possible source of nonlocality in the theory.} You get back what
you put in: if you put the nonlocal collapse, you get nonlocality.
When we only use the local QFT interactions even to describe the
measuring process, we get a fully local theory.

Note that in ref.~\cite{Ballentine70} the EPR theorem led to the
conclusion that the quantum theory had to be completed: the single
system had to have precise values of anticommuting observables
like momenta and positions, which played the role of hidden
variables. As I have commented in the introduction, this was
precisely the kind of solution that Einstein tended to favor,
however (at least in its radical version) it is ruled out by
Bell's theorem. Here, after removing EPR theorem, we do not need
to introduce hidden variables. Ironically, we have come to a
similar conclusion as EPR (the state vector does not describe the
single system), although now we do not need to provide a more
complete description! This fact may have important philosophical
consequences.

It has been claimed that in statistical interpretations which do
not introduce hidden variables ``all the systems that have the
same wave function are identical, since nothing differentiates
them"~\cite{dEspagnat}. Such a criticism would be correct if we
did attribute the wave function to the single system. In a purely
probabilistic theory, the only conclusion that could be reached
with this kind of consideration would be that the minimal
statistical interpretation does not explain \emph{why} the
measurements give sharp, different values, although it allows for
this. On the other hand, the unique known kind of interpretations
that would give a real ``explanation" would be those based on
hidden variables, which have problems with Bell's theorem (an
exception is Bohm's theory, which I will not consider here since
it is non-relativistic from the beginning and it has not been
implemented to an interpretation of QFT). May be that we will
never solve this mystery, which might reach the limits of science,
although I hope that some progress will be provided with a
possible Theory of Everything. For the moment, QFT with its
statistical interpretation is not an \emph{explanation,} but at
least it is a consistent (and local) \emph{description}.

\section{The EPR correlations in QFT}

Although we have already seen that QFT with the statistical
interpretation is a local theory, it is interesting to discuss how
this fact is shown in the EPR correlations that it predicts, whose
supposed nonlocality is thought to be proven by the generalized
Bell's Theorem (the only remaining argument after the removal of
the EPR+Bell argument).

Let us assume for simplicity that in an EPR-Bohm experiment the data
analysis only records the ``coincident events" with a particle A
appearing in detector $O_A$ and a (say different) particle B appearing
in $O_B$ (but all the following discussion can be easily generalized
to the case of also considering the coincident events when B is caught
by $O_A$ and A is caught by $O_B$, and to the case of A and B being
identical, indistinguishable particles). I will call ${\cal E}$ the
statistical ensemble of such coincident events, that have to be
selected by two local measurements on both particles A and B.  Suppose
that $O_A$ measures the component $\vec S(A)\cdot\vec a$ of the spin
of A, and $O_B$ measures the component $\vec S(B)\cdot\vec b$ of the
spin of B. Let $s_a=\pm\hbar/2$ and $s_b=\pm\hbar/2$ indicate the
corresponding eigenvalues (i.e.\ the possible results of a
measurement) of these spin projections.\footnote{More precisely, we
understand that a basis of Dirac spinors has to be used.}

Now, the statistics over the ensemble ${\cal E}$ allows for the
experimental evaluation of the spin correlations $\langle\vec S(A)
\cdot\vec a \,\vec S(B)\cdot\vec b\rangle$. In QFT, such a correlation
function has to be computed from the rate $\Gamma(s_a,s_b)$ for those
coincident events where particle A and B are found with definite
values $s_a$ and $s_b$ of their spin components. To give an example, I
will assume that A and B are created in the decay process of a
particle X (e.g.\ we may consider the decay of the possible Higgs
boson into an electron-positron pair, that can be calculated in a
straightforward way in the Standard Model), but the whole discussion
which follows can be easily generalized e.g.\ to the case when A and B
arise from a scattering process. I will compute the correlation at the
lowest order, neglecting the rates for the creation of additional
photons in coincidence with A and B. In fact, although the effect of
these additional photons was important for removing the EPR
incompleteness argument, the contribution of the diagrams involving
additional photons to the spin correlations is small~\cite{pureprp},
due to the fact that their rates are suppressed by powers of the fine
structure constant and by a phase space factor depending on the small
solid angles $\Omega_A$ and $\Omega_B$ intercepted by the detectors as
seen from the production point (as far as we consider only the
ensemble ${\cal E}$). A good approximation for the relevant QFT rate
will be
\begin{equation}
\Gamma(s_a,s_b) \simeq \frac{c^2}{2(2\pi) ^2\hbar
M_X}\int_{\Omega_A}\frac{d^3\vec
p_A}{2E_A}\int_{\Omega_B}\frac{d^3\vec p_B}{2E_B} \vert {\cal M} (\vec
p_A,s_a,\vec p_B,s_b)\vert^2\delta^4(p_A+p_B-p_X)
\label{eq:2}
\end{equation}
where the variables of integration $p_A=(E_A/c,\vec p_A)$,
$p_B=(E_B/c,\vec p_B)$ correspond to the four momenta of the two
particles A and B, and $M_X$ is the mass of the decaying particle that
produces them (see e.g.\ eq. (A.57) of ref.~\cite{Peskin}).  On the
other hand, ${\cal M} (\vec p_A,s_a,\vec p_B,s_b)$ is the Feynman
amplitude for the process. Assuming that the two detectors are placed
in exactly opposite directions and intercept similar solid angles, we
can perform the integrals and get
\begin{equation}
\Gamma(s_a,s_b) \simeq \frac{\Omega_A}{4\pi}\frac{\vert \vec
p_A\vert}{8\pi\hbar c M_X^2} \vert {\cal M} (\vec p_A,s_a,-\vec
p_A,s_b)\vert^2\,, \label{eq:3}
\end{equation}
where now 
$$
\vert \vec p_A\vert =
cM_X^{-1}\sqrt{M_X^4+M_A^4+M_B^4-2M_X^2M_A^2-2M_X^2M_B^2-2M_A^2M_B^2}
$$
is the modulus of the momentum, as given by energy conservation (here
we have chosen the rest frame of the decaying particle).

Restricting ourselves to the statistical ensemble ${\cal E}$, we can
define the probability of obtaining $s_a$ and $s_b$ by dividing the
rate in eq.~(\ref{eq:3}) by the total rate corresponding to this
statistical ensemble, i.e.\ 
\begin{equation} 
{\cal P}(s_a,s_b) \equiv \frac{\Gamma(s_a,s_b)}{
\sum_{s'_a=\pm\hbar/2}\sum_{s'_b=\pm\hbar/2}\Gamma(s'_a,s'_b)}
\simeq\frac{\vert {\cal M} (\vec p_A,s_a,-\vec p_A,s_b)\vert^2}
{\sum_{s'_a=\pm\hbar/2}\sum_{s'_b=\pm\hbar/2}\vert {\cal M} (\vec
p_A,s'_a,-\vec p_A,s'_b)\vert^2}\,. \label{eq:4}
\end{equation}

Now, in QFT the Feynman amplitude ${\cal M}$ is invariant under the
full Lorentz group (transforming the momenta and spin/helicities as
shown in ref.~\cite{WeinbookI}). In particular, it is invariant under
rotations, thus it conserves angular momentum. In the usual EPR-Bohm
experiment we assume that the decaying system $M_X$ has zero angular
momentum, thus A and B are created in a spin-singlet state (remember
that we are neglecting any possible additional particles in this
computation).  Therefore, independently of the explicit form of the
amplitude ${\cal M}$, that depends on the particular process under
consideration, we get
\begin{equation}
{\cal P}(+_a,+_b)={\cal
P}(-_a,-_b)\simeq\frac12\sin^2\left(\frac\theta2\right),\qquad 
{\cal P}(+_a,-_b)={\cal P}(-_a,+_b)\simeq\frac12
\cos^2\left(\frac\theta2\right),
\label{eq:5}
\end{equation}
where $\theta$ is the angle between the orientations $\vec a$ and
$\vec b$ of the spin-measuring devices, and I have indicated
$\pm\hbar/2$ with the sign $\pm$. We can now evaluate the
spin-spin correlation (for the ensemble ${\cal E}$) as
\begin{equation}
\langle s_a s_b \rangle =
\sum_{s_a=\pm\hbar/2}\sum_{s_b=\pm\hbar/2}{\cal P}(s_a,s_b) s_a
s_b \simeq -\frac{\hbar^2}{4}\cos\theta\,. \label{eq:6}
\end{equation}
This is the same result obtained in the usual QM
treatment~\cite{BJ,Laloe}. There are however two important
differences.  Firstly, now we know that it needs two local
measurements on both A and B, and that it is merely an
approximation. Secondly, we see that this result arises from
eq.~(\ref{eq:4}), which only depends on the invariant Feynman
amplitude ${\cal M}$. Now, in the computation of the Feynman
amplitudes all the conservation laws have a local
origin~\cite{WeinbookI,WeinbookII}, and in particular the global
angular momentum conservation is a consequence of the local
conservation, as we expected from the discussion of the previous
section.

This is also reflected by the fact that the correlation does not
depend on the distance: it is the same for large and small
distances. It is true that one can then change the axis $\vec a$ and
$\vec b$ of the distant detectors, but the correlation that is
computed with the new axis could also be computed for small
distances. The correlation has then to be created in the production
process; although it depends nontrivially on the experimental
arrangement in $O_A$ and $O_B$, it does not corresponded to any
influence at a distance.

In fact, eq.~(\ref{eq:5}) satisfies the only real locality
conditions that are imposed by Special Relativity, according to
Ballentine and Jarrett~\cite{BaJa87}:
\begin{equation}
\sum_{s_b}{\cal P}(s_a,s_b)=\frac12\,, \quad {\rm for} \quad
s_a=\pm\frac\hbar2\,; \qquad 
\sum_{s_a}{\cal P}(s_a,s_b)=\frac12\,,
\quad {\rm for} \quad s_b=\pm\frac\hbar2\,, \label{eq:7}
\end{equation}
valid for whatever choice of orientations $\vec a$ and $\vec b$ of the
detectors. In particular, detector $O_A$ observes the same
distribution of probability independently of the settings of the
distant detector $O_B$. This is also sufficient to prevent the
possibility of any superluminal\pagebreak[3] communication. On the other hand,
eq.~(\ref{eq:6}) coincides with the usual QM result, that agrees with
the actual experiments and violates Bell's inequalities. As Ballentine
and Jarrett have argued, this means that it is not possible to find a
set of parameters $\lambda$, carrying the information on the state and
the initial conditions, so that the probability can be factorized~as
\begin{equation}
{\cal P}(s_a,s_b,\lambda)=\sum_{s'_b}{\cal P}(s_a,s'_b,\lambda)
\sum_{s'_a}{\cal P}(s'_a,s_b,\lambda)\,,
\end{equation}
where the dependence on $\lambda$ in the probabilities corresponds
to partitioning the ensemble of the events in subensembles that
have equal conditions $\lambda$. The impossibility of finding such
a decomposition implies that the measurement on A gives additional
information about the measurement on B, with respect to the
information that can be contained in the state preparation.
Ballantine and Jarrett, implicitly using a statistical
interpretation, have proved that this fact should not be
interpreted as a sign of nonlocality, but that it only proves that
the quantum theory is less predictive than it would be using the
state vector to describe the single system. This reduction of
predictivity is unavoidable as far as we limit the description
only to the statistics on the ensemble of copies of the system.

A complete discussion can be found in ref.~\cite{BaJa87}. Here, I will
merely point out that the measurement on A unavoidably gives an
information for the measurement on B, and this information depends on
the orientation $\vec a$ of the apparatus $O_A$. In fact, let me call
${\cal E}(s_a,s_b)$ the partition of the total ensemble ${\cal E}$ of
the coincident events, where for instance ${\cal E}(+_a,+_b)$ is the
ensemble of the events that have given the results $s_a=+\hbar/2$,
$s_b=+\hbar/2$ after the measurement.  This partition depends on the
choice of the orientations $\vec a$ and $\vec b$ of the apparatuses,
and changes if we modify one of these orientations. Since the results
of the single measurements are not predicted by the theory, a
partition can be obtained only \emph{after} actually performing the
measurements. The probabilities ${\cal P}(s_a,s_b)$ can be computed by
counting the number of events that belong to the different
subensembles ${\cal E}(s_a,s_b)$ of the partition of ${\cal E}$ and
dividing by the total number of events in ${\cal E}$ (of course, this
computation would give the correct probabilities only in the limit
where ${\cal E}$ counts infinite events). It is clear that these
probabilities depend on the partition that is chosen, in other words
they depend nontrivially (and in a ``nonseparable" way) on the
orientations of the two apparatuses and on the results of the single
measurements, without implying any nonlocality.\footnote{In Ballentine
and Jarrett's paper~\cite{BaJa87}, the final considerations on the EPR
incompleteness argument should be updated. Now, QFT satisfies the EPR
criterion of completeness, although it allows for \emph{less} reality
than a classically complete theory. On the other hand, QFT does not
predict anything on the single event, and it is ``predictively
incomplete"~\cite{BaJa87}.} On the other hand, although it cannot be
generalized to prove nonlocality, Bell's Theorem may remain an
important argument against local determinism~\cite{BaJa87}.

\section{Conclusions} 

In a previous paper, I have shown that the Standard Model of Particle
Physics prevents the production of states having a definite number of
particles, contradicting a basic assumption of the EPR argument. Here,
this result has been used to remove one of the supposed proofs of
nonlocality as well, that based on the EPR argument and on the
original Bell's Theorem.  The great uncertainty of QFT also provides a
hint for a (minimal) statistical interpretation that renounces
describing the single events. Such an interpretation does not assume
any global collapse of the state vector and solves the ``measurement
problem" in a natural way, since it allows for describing the
measurement as a physical quantum process, in agreement with recent
results~\cite{Zeno} and with the experience of particle
physicists. This also allows for a complete recovery of locality and
causality {\it without introducing hidden variables.} I have then
shown how the EPR correlations should be computed in QFT using Lorentz
invariant Feynman amplitudes, and recalled the argument by Ballentine
and Jarrett against the wild generalizations of Bell's Theorem. The
latter theorem may still be used in its original formulation to rule
out local determinism, but it cannot prevent the probabilistic QFT to
be a local theory. We have no guarantee that a better understanding of
the quantum correlations will ever be found, although it can be hoped
that this will eventually be achieved in a possible ultimate Theory of
Everything. Further research in this direction will be most welcome
(e.g.\ following the ideas given in refs.~\cite{Porto,Czachor}). In
any case, I think that locality and Lorentz symmetry will not be so
badly violated as to allow for whatever instantaneous effect between
very distant objects, since QFT will presumably remain a good
approximation in the ``low energy" regime (corresponding precisely to
the long distance behavior). On the other hand, these results can be
expected to be maintained or even strengthened in a possible quantum
gravity, since General Relativity has been argued to require a local
interpretation of the correlations~\cite{Mensky}.

\acknowledgments 

I thank Humberto Michinel for very useful and stimulating discussions,
and Rebecca Ramanathan and Ruth Garc\'\i a Fern\'andez for help.

\end{document}